\documentclass[showpacs,preprintnumbers,amsmath,amssymb]{revtex4}

\usepackage{graphicx}
\usepackage{dcolumn}
\usepackage{bm}
\usepackage{url}
\usepackage{epsfig}
\usepackage{multirow}

\newcommand{\be}{\begin{equation}}
\newcommand{\ba}{\begin{eqnarray}}
\newcommand{\ee}{\end{equation}}
\newcommand{\ea}{\end{eqnarray}}
\newcommand{\fr}{\frac}
\newcommand{\oo}{\omega_{0}}
\newcommand{\oa}{\omega_{a}}

\newcommand{\PRD}{Phys.\ Rev.\ D}
\newcommand{\MNRAS}{Mon.\ Not.\ Roy.\ Astron.\ Soc.}

\begin{document}

\title{Constraints on the dark energy using multiple observations : snare of principal component analysis}


\author{Seokcheon Lee}
\email{skylee@kias.re.kr}
\affiliation{School of Physics, Korea Institute for Advanced Study, Heogiro 85, Seoul 130-722, Korea}


\begin{abstract}
We explore snares in determining the equation of state of dark energy ($\omega$) when one uses the so-called principal component analysis for multiple observations. We demonstrated drawbacks of principal component analysis in an earlier paper \cite{10051770}. We used the Hubble parameter data generated from a fiducial model using the so-called Chevallier-Polarski-Linder parameterization. We extend our previous consideration to multiple observations, the Hubble parameter and the luminosity distance. We find that the principal component analysis produces the almost constant $\omega$ even when a fiducial model is a rapidly varying $\omega$. Thus, resolution of dynamical property of $\omega$ through PCA is degraded especially when one fits to several observations.
\end{abstract}


\maketitle

The existence of the energy component with negative pressure ({\it i.e.} dark energy) besides the matter component is one of the alternative elucidations of the current accelerating expansion of the Universe. The lack of enthralling theoretical model enforces us to use its equation of state (EOS) $\omega$ to classify the different dark energy (DE) models. By using the accumulating high precision observational data, the dark energy can be properly investigated. The proper parametrization of $\omega$ is prerequisite for fitting the related parameters to data. One of the model-independent reconstruction of $\omega$ is to approximate it by using the piecewise constant bins \cite{0106079, 0207517}. It is claimed that one can reconstruct the time dependence of $\omega$ and make further model independent studies by using a principal component analysis (PCA) method. This method identifies the directions of data points clustering in the parameter space, and then find a viable dimensional reduction in the parameters with as minimum an information loss as possible \cite{9603021}. However, it is argued that PCA is sensitive to both the order and the basis set choice \cite{09053383}. Our conclusion is against this argument. We found that PCA produces the almost same results independent of the order of principal components as shown in Ref. \cite{10051770}. We will also show that PCA gives the almost identical results when we change the basis set. One of the main reason for using PCA method is known as to determine whether the dark energy density is evolving with time or not. However, it is shown that the time varying $\omega$ can be confounded with the incorrect value of constant one when we use PCA to fit the Hubble parameter $H(z)$ data generated from a fiducial model \cite{10051770}. We adopt the so-called  Chevallier-Polarski-Linder (CPL) parameterization $\omega = \omega_{0} + \omega_{a} \fr{z}{1+z}$ \cite{0009008,0208512} as a fiducial DE model. In addition to the Hubble parameter, we use the luminosity distance $D_{L}(z)$ data generated from a fiducial model to investigate the PCA method.

One divides the redshift range of the survey ($z=0, z_{\rm{max}}$) into $N$ bins of not necessarily equal widths $\Delta z_i$ ($i = 1, \cdots , N$), where $\sum_i \Delta z_i = z_{\rm{max}}$. Then, a set of $N$ values of observations of possibly correlated variables can be orthogonally transformed into a set of $j$ values of uncorrelated variables so-called principal components without losing the original information much \cite{Pearson, 9603021}. The dark energy is parameterized in terms of $\omega(z)$, which is defined to be a constant in each redshift bin, with a value $\omega_{i}$ in $i$th bin. For the piecewise constant $\omega(z)$, the energy density of the dark energy for $z$ in bin $j$ evolves as \be \rho_{\rm{DE}}(z) = \rho_{\rm{DE}}(z=0) \Biggl(\fr{1+z}{1+z_j} \Biggr)^{3(1+\omega_j)} \prod_{i=1}^{j-1} \Biggl(\fr{1+z_{i+1}}{1+z_i} \Biggr)^{3(1+\omega_i)} \, , \label{rhode} \ee where $z_i$ is the lower redshift bound of the $i$th bin and $\omega_i$ is the fiducial value of the EOS in that bin. Again, we emphasize that $\omega_i$s have only $j-1$ degree of freedom because $\rho_{DE}(z)$ should be equal to $\rho_{DE}(z=0)$ when $z=0$ \cite{10051770}. From the above equation (\ref{rhode}), this is given by \be \omega_j = -1 + \Biggl( \sum_{i=1}^{j-1} (1+\omega_i) \ln \Bigl[\fr{1+z_{i+1}}{1+z_i} \Bigr] \Biggr) \Biggl/ \Bigl( \ln [ 1+z_j ] \Bigr) \, . \label{omegaj} \ee Thus, $\omega_{j}$ is determined by other parameters $\omega_i$, $z_i$, and $z_{j}$.

\begin{center}
    \begin{table}
    \begin{tabular}{ | c | c | c | c | c | c | c | c | c | c | c | c |  }
    \hline
      $\oo$ & $\oa$ & $z_{i}$ & \multicolumn{3}{|c|}{$\omega_{i}$} & \multicolumn{3}{|c|}{$\sigma_{i}$} & \multicolumn{3}{|c|}{$\chi_{{\rm min}}^2$} \\ \hline
            &       &         & $H$ & $D_L$ & $H+D_L$ & $H$ & $D_L$ & $H+D_L$ & $H$ & $D_L$ & $H+D_L$ \\ \cline{4-12}
            &       & $0.1$   & $-0.995$ & $-1.011$ & $-1.009$ &  $0.07$  & $0.01$ & $0.04$ & $$ & $$ &           \\ \cline{3-9}
            &       & $0.4$   & $-0.935$ & $-0.984$ & $-0.982$ &  $0.08$  & $0.01$ & $0.04$ & $$ & $$ &           \\ \cline{3-9}
      $-1.1$&$0.5$  & $0.8$   & $-0.874$ & $-0.977$ & $-0.975$ &  $0.12$  & $0.01$ & $0.14$ & $0.11$ & $1.14$ & $2.34$ \\ \cline{3-9}
            &       & $1.25$  & $-0.840$ & $-0.981$ & $-0.979$ &  $0.25$  & $0.01$ & $0.31$ & $$ & $$ &           \\ \cline{3-9}
            &       & $1.6$   & $-0.928$ & $-0.990$ & $-0.988$ &  $$  & $$ & $$ & $$ & $$ &           \\ \hline
            &       & $0.1$   & $-1.217$ & $-1.204$ & $-1.205$ &  $0.08$  & $0.01$ & $0.01$ & $$ & $$ &           \\ \cline{3-9}
            &       & $0.4$   & $-1.196$ & $-1.162$ & $-1.163$ &  $0.10$  & $0.01$ & $0.01$ & $$ & $$ &           \\ \cline{3-9}
      $-1.1$&$-0.3$ & $0.8$   & $-1.225$ & $-1.162$ & $-1.163$ &  $0.18$  & $0.01$ & $0.01$ & $0.33$ & $12.85$ & $13.38$ \\ \cline{3-9}
            &       & $1.25$  & $-1.235$ & $-1.158$ & $-1.159$ &  $0.41$  & $0.02$ & $0.02$ & $$ & $$ &           \\ \cline{3-9}
            &       & $1.6$   & $-1.194$ & $-1.156$ & $-1.157$ &  $$  & $$ & $$ & $$ & $$ &           \\ \hline
            &       & $0.2$   & $-0.980$ & $-0.999$ & $-0.997$ &  $0.08$  & $0.01$ & $0.01$ & $$ & $$ &           \\ \cline{3-9}
            &       & $0.5$   & $-0.905$ & $-0.974$ & $-0.972$ &  $0.09$  & $0.01$ & $0.01$ & $$ & $$ &           \\ \cline{3-9}
      $-1.1$&$0.5$  & $1.0$   & $-0.856$ & $-0.970$ & $-0.968$ &  $0.15$  & $0.01$ & $0.01$ & $0.10$ & $1.88$ & $2.92$ \\ \cline{3-9}
            &       & $1.4$   & $-0.816$ & $-0.975$ & $-0.973$ &  $0.43$  & $0.02$ & $0.02$ & $$ & $$ &           \\ \cline{3-9}
            &       & $1.6$   & $-0.924$ & $-0.984$ & $-0.983$ &  $$  & $$ & $$ & $$ & $$ &           \\ \hline
    \end{tabular}
    \caption{Parameter values of fiducial models are shown in the first two columns. $z_{i}$ is the uncorrelated bin and there are three $\omega_{i}$s at each bin obtained from the minimum $\chi^2$ fitting by using the generated $H(z)$, $D_{L}(z)$, and both of them. $\sigma_{i}$s are the $1$-$\sigma$ errors of $\omega_i$s for corresponding fittings.}
    \label{table1}
    \end{table}
\end{center}

\begin{center}
\begin{figure}
\vspace{1.5cm} \centerline{ \psfig{file=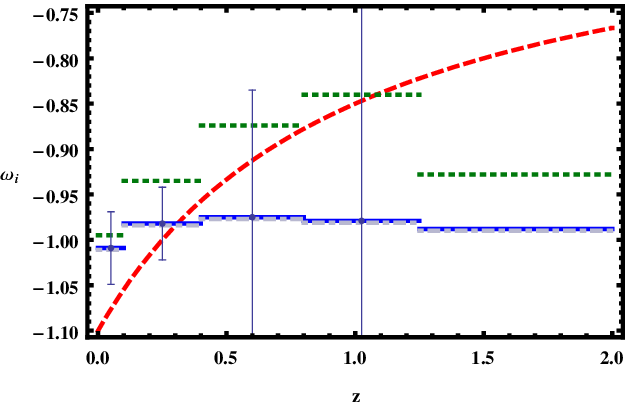, width=6.5cm}
\psfig{file=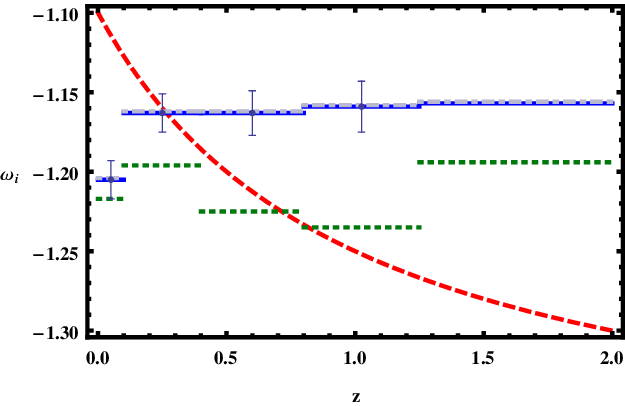, width=6.5cm}
} \vspace{-0.1cm} \caption{
Comparison between a fiducial $\omega$ and $\omega_{i}$s obtained from the different observations.
a) The fiducial model is $\omega = -1.1 + 0.5 \fr{z}{1+z}$ (dashed) and the obtained values of $\omega_{i}$s from PCA with $H$ data (dotted), $D_{L}$ data (dot-dashed), and $H + D_{L}$ data (solid). Error bars are obtained from the analysis of $H + D_{L}$ data. b) The fiducial model is $\omega = -1.1 - 0.3 \fr{z}{1+z}$ (dashed) and the obtained values of $\omega_{i}$s from PCA with $H$ data (dotted), $D_{L}$ data (dot-dashed), and $H + D_{L}$ data (solid).} \label{fig1}
\end{figure}
\end{center}

\begin{center}
\begin{figure}
\vspace{1.5cm} \centerline{ \psfig{file=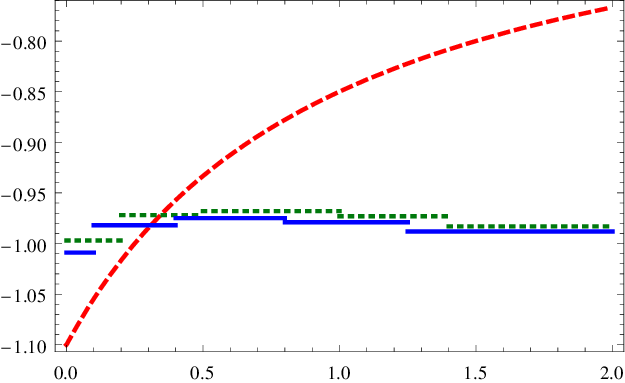, width=8cm}
} \vspace{-0.1cm} \caption{
Comparison of the PCA results for the two different sets of principal components from a fiducial model ($\oo, \oa$) = ($-1.1, 0.5$) (dashed). $zi$ = ($0.1, 0.4, 0.8, 1.25, 1.6$) (solid), ($0.2, 0.5, 1.0, 1.4, 1.6$) (dotted).
} \label{fig2}
\end{figure}
\end{center}

In what follows, we use equally binned both the $40$ Hubble parameter $H(z)$ data and the $2000$ luminosity distance $D_{L}(z)$ data to cover the redshift range $0 < z \le 2$ generated from the fiducial models given by CPL parametrization. We assume that the errors on both measurements as $5$ \% and both the present energy density contrast of the matter $\Omega_{m}^{0}$ and the present Hubble parameter $H_0$ are exactly known values. We also consider only the flat universe. We perform the simple $\chi^2$ tests to determine the best fit values of $\omega_{i}$ and $1$-$\sigma$ error which is obtained from the covariance matrix. By adding the luminosity distance in the $\chi^2$ test, we obtain the more confounded results than the one without it. This is expected due to the multi-integral of $\omega$ in it. We show the results for the two different models in table \ref{table1}. In the first case, a fiducial model is ($\oo, \oa$) = ($-1.1, 0.5$). We perform $\chi^2$ test with both $H(z)$ and $D_{L}(z)$ data generated from this fiducial model. The fiducial model of the second case is ($\oo, \oa$) = ($-1.1, 0.3$). We also investigate the PCA result of the first model with the different bins. We explain the details of the data with Fig. \ref{fig1} and Fig. \ref{fig2} later. We choose the uncorrelated bins as $z_i = (0.1, 0.4, 0.8, 1.25, 1.6$) in the first two cases and $z_i = (0.2, 0.5, 1.0, 1.4, 1.6$) in the third case. As we show before, there are only $4$ degree of freedom in this case. Thus, the value of $\omega_5$ at $z = 1.6$ is derived from the other $\omega_i$, $z_{i}$, and $z_5$ values by using Eq. (\ref{omegaj}).

We find several snares in PCA when we try to fit both $H$ and $D_L$ data. Firstly, if we exclude the annoying value $\omega_{5}$, then the dynamical property of $\omega$ might be detected with PCA from $H$ data as shown in the first fiducial model (dashed line) even though the $\omega_{i}$ values (dotted lines) depart from the values of the fiducial model. However, PCA produces the confounded results when we fit $\omega_i$ to $D_{L}$ data (dot-dashed lines). The result can be interpreted as $\omega = -1$ for the entire region of $z$ within $1$-$\sigma$ error. We reach to the same conclusion when we fit $\omega_i$ to both $H$ and $D_L$ (solid lines). This is shown in the left panel of Fig. \ref{fig1}. The fiducial model is $\omega = -1.1 + 0.5 \fr{z}{1+z}$  but the result from PCA is consistent with $\omega = -1$. Thus, PCA with multiple observations degrades the resolution of $\omega$. One remark is that one might misinterpret that PCA is at least good for fitting $H$ data to reveal the time variation of $\omega$. We showed that even the $\omega$ obtained from the $H$ fitting shows the discrepancy with the fiducial ones \cite{10051770}. Secondly, PCA produces the smaller $\omega_{1}$ value than the fiducial one when $\oa$ is negative. This is similar to the problem of CPL parameterization because it might mislead the property of the fiducial model. It is well known that the negative $\oa$ is harder to be detected compared to the positive one \cite{0112526}. In the right panel of Fig. \ref{fig1}, $\omega_{i}$s increase as $z$ does even though the fiducial model $\omega = -1.1 - 0.3 \fr{z}{1+z}$ decreases as $z$ increases. The result of PCA in this case also can be mislead to the constant $\omega$.  Thus, PCA method produces the totally different behavior of $\omega$ when $\oa$ is non-negligible independent of the sign of it. We can compare this result with one in Ref. \cite{09083186}. Even though the result in the mentioned reference seems to be consistent with the cosmological constant, there still can be the viable time varying DE models which can mimic $\Lambda$. This impedes any proper interpretation of the result obtained from PCA. We check that PCA method can give the reliable result only when $\omega$ is almost constant.

We show the PCA results from the two different sets of principal components in Fig. \ref{fig2}. We show the results using the $H + D_{L}$ data only. The dashed line is the fiducial model  $\omega = -1.1 + 0.5 \fr{z}{1+z}$ to generate the $H$ and the $D_{L}$ data. The solid lines depict $\omega_{i}$s obtained from binning $z$ as ($0.1, 0.4, 0.8, 1.25, 1.6$). The dotted line correspond to the case when $z$ is divided by ($0.2, 0.5, 1.0, 1.4, 1.6$). The results from two different PCs are almost identical.

Even though, PCA is the most model independent method for probing DE, we demonstrate that PCA method may mislead to the property of dark energy independent of the dynamics of the fiducial $\omega$. When one use PCA to several observations, the resolution of $\omega$ is degraded. The principal component is automatically decided when the data is given and our conclusion is independent of PCs. PCA is adequate only when $\omega$ is a constant. Thus, we may need to check both the model dependent $\omega$ parametrization and PCA method to investigate the DE properly.


\end{document}